\documentclass[conference]{IEEEtran}
\IEEEoverridecommandlockouts

\usepackage{cite}
\usepackage{amsmath,amssymb,amsfonts}
\usepackage{algorithmic}
\usepackage{graphicx}
\usepackage{textcomp}
\usepackage{xcolor}
\usepackage{romannum}
\usepackage{multirow}
\usepackage{arydshln}
\usepackage{booktabs}
\usepackage{subfigure}

\def\BibTeX{{\rm B\kern-.05em{\sc i\kern-.025em b}\kern-.08em
    T\kern-.1667em\lower.7ex\hbox{E}\kern-.125emX}}
\begin{document}
\title{Mamba for Streaming ASR Combined with Unimodal Aggregation}

\author{\IEEEauthorblockN{Ying Fang$^{1, 2}$,  Xiaofei Li$^{2,3*}$\thanks{* Corresponding author.}}
\IEEEauthorblockA{\textit{$^{1}$ Zhejiang University, China} \\
\textit{$^{2}$ School of Engineering, Westlake University, China}\\
\textit{$^{3}$ Institute of Advanced Technology, Westlake Institute for Advanced Study, China}\\
\{fangying, lixiaofei\}@westlake.edu.cn}
}

\maketitle



%
\begin{abstract}
This paper works on streaming automatic speech recognition (ASR). Mamba, a recently proposed state space model, has demonstrated the ability to match or surpass Transformers in various tasks while benefiting from a linear complexity advantage. We explore the efficiency of Mamba encoder for streaming ASR and propose an associated lookahead mechanism for leveraging controllable future information. 
Additionally, a streaming-style unimodal aggregation (UMA) method is implemented, which automatically detects token activity and streamingly triggers token output, and meanwhile aggregates feature frames for better learning token representation. 
Based on UMA, an early termination (ET) method is proposed to further reduce recognition latency. Experiments conducted on two Mandarin Chinese datasets demonstrate that the proposed model achieves competitive ASR performance in terms of both recognition accuracy and latency. Code will be open-sourced \footnote{https://github.com/Audio-WestlakeU/UMA-ASR}.
\end{abstract}

\section{Introduction}
\label{sec:intro}
Streaming automatic speech recognition (ASR) has a wide range of real-life applications and has greatly developed in recent years. Due to the misaligned input feature and output token sequences, one core difficulty for streaming ASR is detecting each token's endpoint in real time and immediately outputting the token. End-to-end models, such as connectionist temporal classification (CTC) \cite{graves2006connectionist} and neural transducers \cite{zhang2020transformer}, have become the dominant approaches in streaming systems, where endpoint detection is trained and conducted implicitly and token outputs are indicated with spike probabilities. Typically, these models employ a Transformer-based encoder or its variants, such as the Conformer \cite{gulati2020conformer}. To enable streaming inference with these encoders, two primary strategies are utilized: (\romannum{1}) implementing causal or chunk-masked self-attention \cite{zhang2020unified,strimel2023lookahead} and causal convolutional neural networks (CNNs) to restrict the encoder from accessing 
future frames; (\romannum{2}) adopting block processing \cite{tsunoo2019transformer}, where each input chunk comprises a fixed number of past, current, and future frames, allowing interaction only within the same chunk. 
 CTC-based streaming ASR faces the challenge of delayed emission of non-blank tokens. To address this issue, several regularization methods have been proposed. FastEmit \cite{yu2021fastemit}, Peak-first CTC \cite{10095377}, and Delay-Penalty \cite{kang2023delay} incorporate penalty terms into the loss function to encourage earlier emission of non-blank tokens. TrimTail \cite{10097012} addresses the emission latency by trimming the trailing frames of the spectrogram. All these methods are applied during training and achieve a trade-off between latency and accuracy by adjusting hyperparameters.


Exploring streaming strategies for triggering the auto-regressive decoder is also one of the research hotspots.
The triggered attention system \cite{moritz2020streaming} uses CTC output spikes to trigger the decoder.
Continuous integrate-and-fire (CIF) \cite{dong2020cif} and cumulative attention (CA)\cite{li2022transformer,10094800} accumulate acoustic features 
until the accumulated confidence reaches a pre-designed threshold.
Inspired by the success of decoder-only large language model, \cite{chen2024streaming} investigates the decoder-only streaming ASR by inserting "boundary tokens" into the discrete speech token sequence, relying on the forced alignment obtained by a GMM-HMM model. 

State-space models (SSMs), such as Mamba \cite{gu2023mamba,pmlr-v235-dao24a}, have recently demonstrated comparable or superior performance to Transformers in various tasks. One critical limitation of the self-attention mechanism lies in its quadratic scaling to input sequence length, whereas Mamba exhibits the advantage of linear scaling. For ASR, the input frame rate is approximately 30 frames per second, employing Mamba as the encoder will substantially reduce computational costs as speech utterance lengthens. 
Quite recently, \cite{zhang2024mamba,jiang2024speech} extended Mamba to be bi-directional and used it for offline ASR, which showed certain performance superiorities. 

In this work, within a unimodal aggregation (UMA) framework, we propose a streaming ASR model by exploiting Mamba encoder. UMA was proposed in our previous work \cite{fang2024unimodal} for offline ASR. In UMA, one text token has unimodal weights (namely first monotonically increasing and then decreasing weights) on feature frames that belong to the token. The unimodal weights are derived from encoder features and then used for segmenting and integrating the feature frames. 
Explicitly integrating the feature frames enhances feature quality, thereby improving ASR performance. Additionally, the unimodal weights provide explicit token boundaries, naturally addressing the core difficulty in streaming ASR: endpoint detection for triggering token output. 
This work investigates the efficiency of UMA for streaming ASR. Moreover, based on the unimodal weights, an early termination (ET) method during inference is proposed to further reduce the recognition latency. 
The inherent causal structure of Mamba renders it exceptionally well-suited for streaming ASR. This work investigates the efficiency of Mamba encoder for streaming ASR, and develops an associated lookahead mechanism to better tradeoff recognition accuracy and latency. 
As demonstrated in \cite{huang2014time}, native Chinese speakers require approximately 480 milliseconds to recognize the first syllables in Mandarin disyllabic words, which means certain recognition latency is tolerable in speech interaction. In Transformer, lookahead can be easily realized by allowing self-attention to future frames. In this work, accompanying the Mamba encoder, a simple yet effective convolution-based lookahead mechanism is designed.
Overall, by properly integrating the Mamba encoder, lookahead mechanism and UMA, the proposed streaming ASR model achieves state-of-the-art (SOTA) performance on two Mandarin ASR datasets.   

\begin{figure}[t]
\centering
\centerline{\includegraphics[width=0.98\columnwidth]{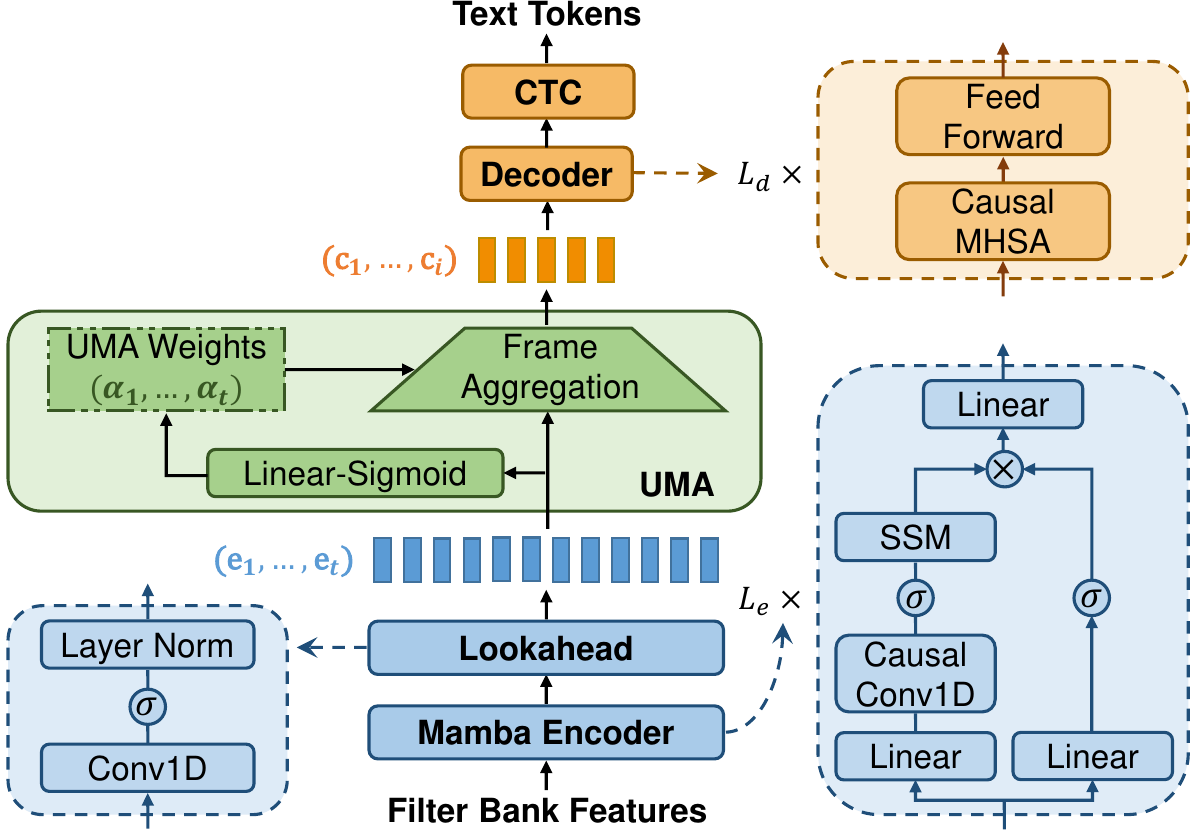}}
\caption{Model architecture. $\sigma$ represents activation layer. Residual connection and normalization are omitted in encoder and decoder.}
\label{fig:mamba}
\vspace{-0.4cm}
\end{figure}

\section{Method}
\label{sec:method}
Fig.~\ref{fig:mamba} shows the architecture of the proposed streaming ASR model, which consists of a Mamba encoder, followed by a convolution look-ahead layer, then a UMA module, and finally a self-attention decoder to output the recognized text. The whole model is trained end-to-end with CTC loss.  
\subsection{Mamba Encoder}
The structured SSMs employ a state-space representation $\mathbf{h} \in \mathbb{R}^{({N}, {T})}$ to model sequence-to-sequence transformation, where ${N}$ and ${T}$ denote state size and sequence length, respectively. Mapping $D$-dimensional input $\mathbf{x} \in \mathbb{R}^{({D}, {T})}$ to output $\mathbf{y} \in \mathbb{R}^{({D}, {T})}$ is conducted by applying the following discrete process independently to each dimensional of $D$: 
\begin{align}
    h_t = \mathbf{A} h_{t-1} + \mathbf{B} x_{t}; \quad y_t = \mathbf{C}^{T}_t h_t    
\end{align}
where $x_{t} \in \mathbb{R}$ and $y_{t} \in \mathbb{R}$ are the input and output sequence of one dimension. The parameters $\mathbf{A} \in \mathbb{R}^{{N}\times{N}}$, $\mathbf{B} \in \mathbb{R}^{{N}\times 1}$, $\mathbf{C} \in \mathbb{R}^{{1}\times {N}}$ are set to be input-dependent in Mamba \cite{gu2023mamba}, empowering SSM to focus on or disregard different information at various sequence positions. Matrix $\mathbf{A}$ is structured to be diagonal for efficient training.
SSM compresses all the historical information into a constant ($N$)-dimensional of state space, and has a linear computational complexity w.r.t the sequence length ${T}$.

As illustrated in Fig.~\ref{fig:mamba}, one Mamba block consists of a selective SSM module, a linear layer and a 1-dimensional causal convolution layer (with a kernel size of 4) before SSM, a linear gate branch alongside the SSM branch and finally a linear layer after the gated SSM branch. The input linear layers expand the model dimension ${D}$ by a expansion factor ${E}$, resulting in the parameters of a single Mamba block being approximately 3${E}{D}^2$. We repeat the Mamba block $L_e$ times to construct the encoder of the proposed streaming ASR model. 
 
\subsection{Convolutional Lookahead Layer}
We propose integrating a simple yet effective lookahead mechanism after the Mamba encoder to leverage information from future frames, and thus to improve the model's recognition accuracy. It comprises a 1-dimensional non-causal convolution layer, a Swish activation layer, and a Layernorm layer. The extent of future frames leveraged by the lookahead mechanism is regulated through the kernel size ${k}$, and simply being $\frac{{k}-1}{2}$ frames.

\subsection{Unimodal Aggregation}
UMA \cite{fang2024unimodal} was proposed for offline non-autoregressive ASR. In this work, we implement the streaming style of UMA.

After the encoder and lookahead module, the speech embedding sequence $\mathbf{e}_t, t=1, \dots,T$ will be segmented and aggregated to the text token level via UMA. The feature frames that belong to one text token have unimodal weights (which first increases and then decreases). The weights $\alpha_t, t=1,\dots,T$ are computed with a Linear-Sigmoid network taking as input the embedding sequence. 
The timestep $t$ satisfying $\alpha_t\le \alpha_{t-1} \ \text{and} \ \alpha_t \le \alpha_{t+1}$ is defined as UMA valley, while satisfying  $\alpha_t\ge \alpha_{t-1} \ \text{and} \ \alpha_t \ge \alpha_{t+1}$ as UMA peak conversely. 
The embedding of frames that have unimodal weights (in between two consecutive UMA valleys $\tau_{i}$ and $\tau_{i+1}$) are aggregated as:
\begin{align}
\label{eq:uma}
\mathbf{c}_i=\frac{\sum_{t=\tau_{i}}^{\tau_{i+1}}\alpha_t \mathbf{e}_t}{\sum_{t=\tau_i}^{\tau_{i+1}}\alpha_t}.
\end{align}
The aggregated frames will be processed by the decoder (presented later) and then output the recognized text. The assertion that feature frames belonging to one text token have unimodal weights is actually our initial assumption, based on which we aggregate frames following Eq.~(\ref{eq:uma}). When training the whole network end-to-end with the CTC loss, the learned aggregation weights indeed agree with our assumption. This agreement ensures the validity of UMA. Fig.~\ref{fig:uma} illustrates an example of UMA for cases with or without lookahead.


\subsection{Self-attention Decoder}
After UMA, the sequence length is reduced to the token level (about one-fifth of the frame-level length). The aggregated embedding sequence is then processed using a decoder that consists of $L_d$ causal self-attention blocks. 

\subsection{Streaming Inference with Early Termination}
\begin{figure}[t]
\centering
\centerline{\includegraphics[width=0.9\columnwidth]{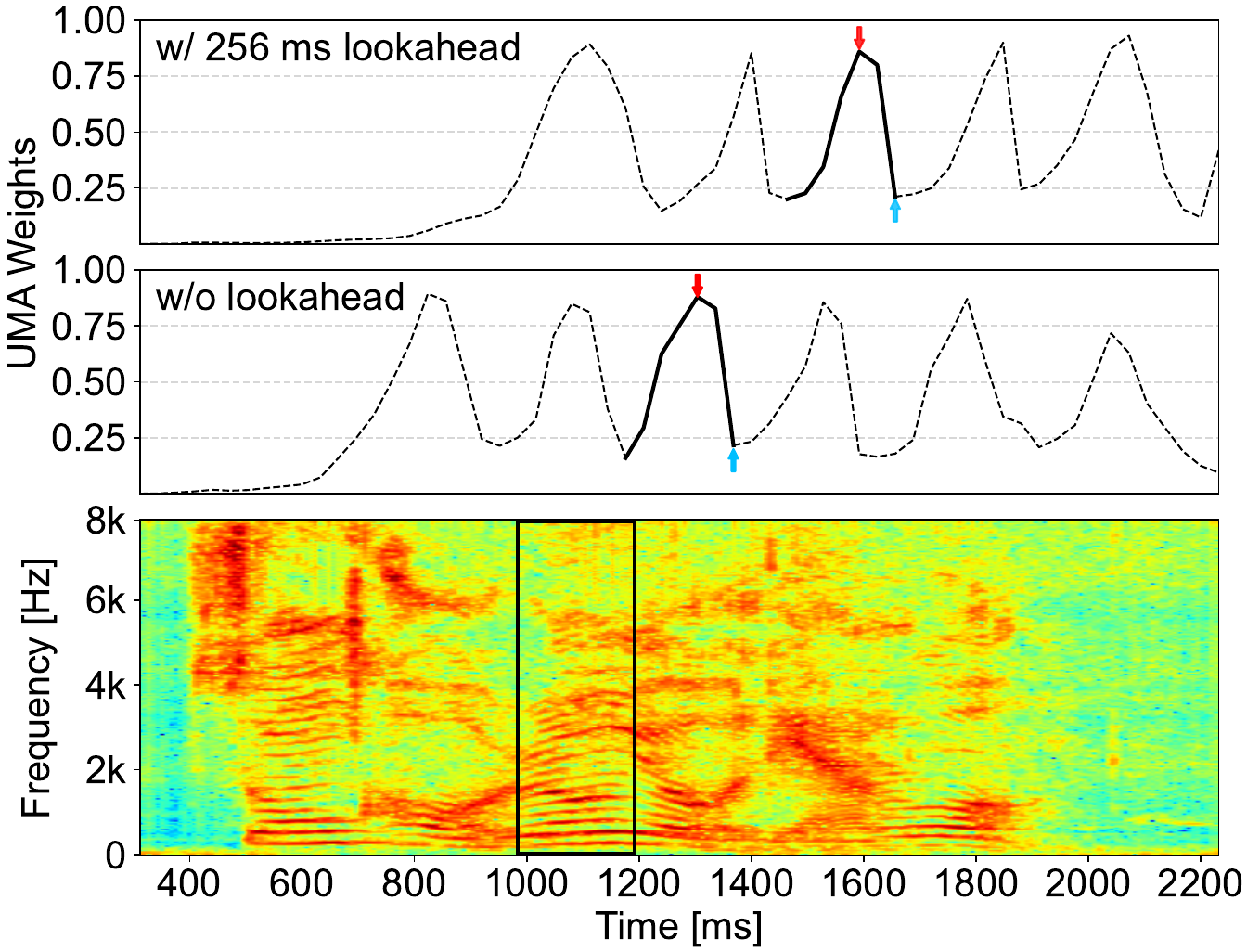}}
\caption{An example of streaming UMA. The spectrogram and UMA weights marked in solid box/line correspond to one same character. 
The blue and red arrows mark a UMA valley and peak, respectively. 
}
\label{fig:uma}
\vspace{-0.4cm}
\end{figure}

During training, all frames are processed in parallel. At inference of streaming ASR, the input speech is processed frame by frame with the encoder, lookahead mechanism and UMA. Once reaching a UMA valley (the endpoint of one text token), the embedding of past frames since the previous UMA valley are aggregated, one time step of decoder is activated and one text token is output. To reduce the recognition latency, we propose an early termination (ET) strategy (only used at inference). Besides outputting a text token when reaching a UMA valley, we give an extra try of outputting a text token when reaching a UMA peak, by aggregating frames from the previous UMA valley to the UMA peak. From Fig.~\ref{fig:uma}, we can see that, from one UMA valley to its succeeding UMA peak, there possibly exists sufficient information for outputting the text token. At the peak, the model could output i) the correct token, then the repeated output at the next valley will be removed and the recognition latency is reduced; ii) a blank token, then the model will correctly output the token until the next valley, resulting in unchanged recognition accuracy and latency; iii) an erroneous token, which brings extra recognition errors. Experiments show that the proposed ET strategy is efficient for reducing recognition latency on average.

\section{Experiments}
\label{sec:experiments}

\subsection{Dataset}

Experiments are conducted on two Mandarin Chinese datasets: AISHELL-1\cite{bu2017aishell} (178 hours) and AISHELL-2\cite{du2018aishell} (1000 hours). As for AISHELL-1 and AISHELL-2 respectively, the test set contains 7176 and 5000 utterances with an average duration of 5.03 and 2.88 seconds, 4,232 and 5,211 Mandarin characters are used as recognition tokens. 

\subsection{Experimental Setup}
Our models are all implemented based on the CTC method within ESPnet\cite{watanabe2018espnet}.

First, we compute 80-dimensional filter bank features using a 32 ms window and an 8 ms shift. Then feature sequence is passed through two causal 2-dimensional convolution layers with a stride of 2, resulting in a frame shift of 32 ms.

\noindent\textbf{Encoder:} Besides the Mamba (https://github.com/state-spaces/mamba) encoder, two other streaming ASR encoders, i.e. causal Transformer and chunk Conformer \cite{zhang2020unified}, are also tested. Causal Transformer is realized by adding triangular mask to attention maps of Transformer encoder. In chunk Conformer, causal CNNs with a kernel size of 15 are used, and the chunk mask is set to 20 (640 ms) at inference. 
The dynamic training method \cite{zhang2020unified} is employed for Conformer training. 
Both the CTC method with or without using the proposed UMA are tested. 
To make the model size of three encoders comparable, the number of encoder blocks for Conformer, Transformer and Mamba are set to 15, 30 and 45 when not using UMA, and to 12, 24 and 36 when using UMA, respectively. Note that an extra decoder (presented later) will be used when using UMA.
For Transformer and Conformer encoders, the model dimension, feedforward dimension and number of heads are set to  256, 2048, and 4 in AISHELL1 experiments and  512, 2048 and 8 in AISHELL2 experiments, respectively. For Mamba encoder, the model dimension, expansion factor and state size are set to 256, 4, 16 in AISHELL1 experiments and  512, 2 and 32 in AISHELL2 experiments, respectively.

\noindent\textbf{Decoder:} A 6-layer causal self-attention decoder is employed when UMA is used. The parameter settings match those of the Transformer encoder for respective datasets.

\noindent\textbf{Optimizer:} We use Adamw optimizer and warmup scheduler. The hyperparameters for learning rate, weight decay, and warmup steps are set to \{0.001, 0.01, 25000\} in AISHELL-1 experiments. For AISHELL-2, we adjust them to \{0.0005, 0.1, 30000\}, and we additionally use a gradient accumulation step of 2. All models are trained using a bath size of 128. Training convergence is determined with validation loss. 

\subsection{Latency Metrics}
ASR accuracy is measured with character error rate (CER). 
To measure the character-level recognition latency, the character boundaries in signals are obtained using the Montreal Forced Aligner tool\cite{mcauliffe2017montreal}, and considered as ground truths. 
For each inference model, the timestamps of token outputs are logged. 
Specifically, for models that utilize chunks, the output time for all tokens within a chunk corresponds to the end time of the chunk. For models using the proposed lookahead mechanism, the convolution of future time is taken into account. 
The recognition latency of each token is calculated by subtracting the ground-truth end time of the corresponding token from the output timestamp. To reflect different delay requirements of streaming ASR in practical applications such as real-time subtitles and voice assistants, three latency measures are computed: (1) \textbf{First Token (FT) Latency}, the recognition latency of the first token in each utterance,  
(2) \textbf{Last Token (LT) Latency}, the latency of the last token in each utterance,  
(3) \textbf{Average (Avg.) Latency}, the average latency of all tokens. 
The average values across all test utterances are reported for each measure, where the worst 10 percent of the tokens are considered outliers and excluded.



\begin{table}[tbp]
\vspace{-0.2cm}
\centering
\caption{Streaming ASR results on AISHELL-1 test set. The Avg. latency values in (\ ) are obtained by correcting the reported values in respective papers with the inherent chunk latency.}
\label{tab:aishell-1}
\setlength{\tabcolsep}{0.4mm}{
\begin{tabular}{l|c|cccc|c}
\hline
\multicolumn{1}{c|}{\multirow{2}{*}{\textbf{Model}}} & \multirow{2}{*}{\textbf{Lookahead}} & \textbf{CER} & \textbf{FT} & \textbf{LT} & \textbf{Avg.} &{\textbf{Params}}  \\ 
 & & (\%) &(ms) &(ms) & (ms) & (M)\\
\hline
Peak-first (3.0) CTC \cite{10095377} & 510 ms   & 6.84 & - & - & (780) & - \\
CA Transformer \cite{10094800} & 1.28 s chunk   & 6.6 & - & - & 624 & 30.2 \\
TrimTail(100) Conformer \cite{10097012} & 640 ms chunk & 6.48 & - & - & (432) & - \\
S3 decoder-only \cite{chen2024streaming} & next token & 6.4 & - & - & - & 70 \\
\hline
\textbf{CTC} &   &  &  &  &  &  \\
Causal Transformer  & 0  & 8.31 & 228 & 140 & 223 & 42.4 \\
Chunk Conformer  & 640 ms chunk & 7.49 & 619 & 418 & 727 & 42.5 \\
Mamba  & 0 & 7.64 & 280 & 176 & 267 & 43.5 \\ 
\hline
\textbf{UMA} &   &  &  &  &  &  \\
Causal Transformer & 0 & 7.08 & 371 & 343 & 369 & 42.5 \\
Chunk Conformer  & 640 ms chunk & 6.04 & 546 & 420 & 642 & 42.5 \\ 
\hdashline
\multirow{2}{*}{Mamba (prop.)} & 0 & 6.59 & 281 & 327 & 271 & \multirow{1}{*}{42.5} \\ 
     & 256 ms & \textbf{5.55} & 605 & 453 & 568 & 43.5 \\ 
\multirow{2}{*}{\ \ with ET }    & 0 & 6.82 & \textbf{212} & \textbf{140} & \textbf{196} & 42.5 \\
    & 256 ms & \textbf{5.55} & 499 & 453 & 494 & 43.5 \\ 
\hline
\end{tabular}
}
\vspace{-0.2cm}
\end{table}

\begin{table}[tbp]
\vspace{-0.2cm}
\centering
\caption{Streaming ASR results on AISHELL-2 iOS test set.}
\label{tab:aishell-2}
\setlength{\tabcolsep}{0.7mm}{
\begin{tabular}{l|c|cccc|cccc|c}
\hline
\multicolumn{1}{c|}{\multirow{2}{*}{\textbf{Model}}} & \multirow{2}{*}{\textbf{Lookahead}} & \textbf{CER} & \textbf{FT} & \textbf{LT} & \textbf{Avg.} &{\textbf{Params}}  \\ 
 & & (\%) &(ms) &(ms) & (ms) & (M)\\
\hline
CIF (chunk-hopping) \cite{dong2020cif} & 2.56 s chunk & \textbf{6.04} & - & - & - & - \\
L5 decoder-only \cite{chen2024streaming} & next token & 7.2 & - & - & - & 310 \\ 
\hline
\textbf{UMA} &   &  &  &  &  &  \\
 Causal Transformer& 0 & 8.45 & 321 & 265 & 299  & 104.8 \\
 Chunk Conformer & 640 ms chunk & 6.98 & 711 & 293 & 496  & 105.0 \\ 
\cdashline{1-7} 
 \multirow{2}{*}{Mamba (prop.)}  & 0 & 7.02 & 314 & 240 & 281   & \multirow{1}{*}{92.3} \\ 
       & 448 ms & \textbf{6.08} & 768 & 335 & 764 & 99.6 \\
\multirow{2}{*}{\ \ with ET } 
  &   0 &7.33 & \textbf{257} & \textbf{179} & \textbf{223} & 92.3  \\
   & 448 ms & 6.25 & 722 & 335 & 699 & 99.6 \\
\hline
\end{tabular}
}
\vspace{-0.4cm}
\end{table}

\begin{figure}[t]
\centering
\subfigure[Performance measures as a function of lookahead]{
    \centering
    \includegraphics[width=0.75\columnwidth]{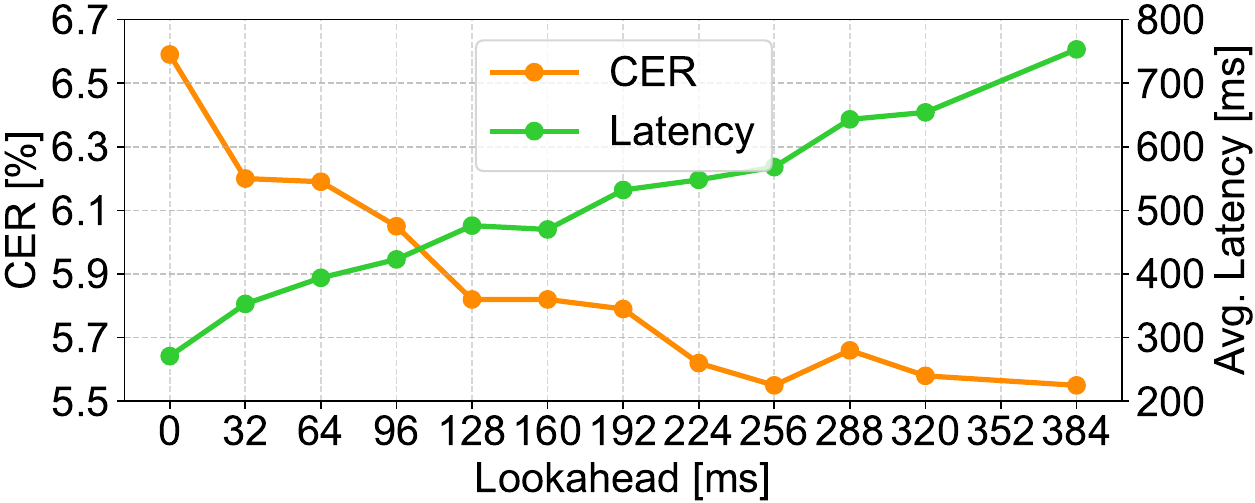}
    \label{fig:aishell1_cer}
}
\subfigure[CER-latency tradeoff with and without ET]{
    \centering
    \includegraphics[width=0.75\columnwidth]{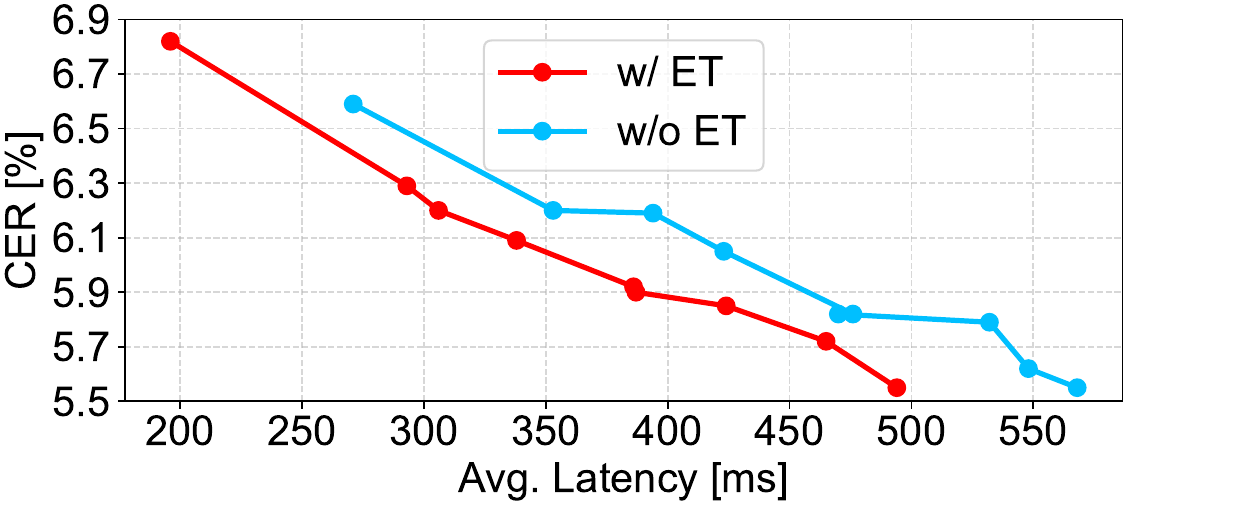}
     \label{fig:aishell1_et}
}
\caption{Experimental results of the lookahead mechanism and ET method with Mamba UMA on AISHELL-1. 
}
\label{fig:aishell1_tradeoff}
\vspace{-0.4cm}
\end{figure}

\subsection{ASR Results}
\label{sec:result}

Table~\ref{tab:aishell-1} and Table~\ref{tab:aishell-2} present the results on AISHELL-1 and AISHELL-2, respectively. 

\textbf{Comparing different encoders.} Comparing the CTC models in Table~\ref{tab:aishell-1}, it is evident that Mamba achieves noticeably lower CER than causal Transformer, and the CER of Mamba is even close to the one of chunk Conformer that leverages much more future information, which demonstrates the clear superiority of Mamba for streaming ASR.   
Although Causal Transformer and Mamba do not apply any specific mechanism for leveraging future information, they have certain recognition latency due to the inherent delay of CTC output. 
Chunk Conformer has a much larger latency, as the tokens within one chunk are output together at the end of the chunk. 
From Table~\ref{tab:aishell-1}, it also can be seen that using UMA significantly improves the performance of all three encoders. By explicitly aggregating feature frames, UMA improves the representation quality of tokens, compared to the implicit information aggregation in CTC. Please refer to the example in Fig.~\ref{fig:uma} to see how UMA works for streaming ASR. 


\textbf{Analysis of the proposed model.} In both Table~\ref{tab:aishell-1} and \ref{tab:aishell-2}, we find that the ASR accuracy of Mamba UMA can be largely improved by applying the lookahead mechanism at a cost of increased latency.   Fig.~\ref{fig:aishell1_cer} illustrates the variation of CER and latency along the increase of lookahead, on AISHELL-1. These results indicate that our proposed lookahead mechanism establishes an effective trade-off between latency and accuracy. The CER reduction converges at about 256 ms of lookahead. Note that, the CER reduction converges at about 448 ms of lookahead for AISHELL-2. 

On the other hand, the proposed ET strategy can effectively reduce recognition latency. As shown in Table~\ref{tab:aishell-1} and \ref{tab:aishell-2}, the average latency can be reduced to as low as 196 ms for AISHELL-1 and 223 ms for AISHELL-2, when lookahead is 0. However, using ET may lead to more recognition errors when erroneous tokens are output at UMA peaks. To evaluate the effectiveness of ET, Fig.~\ref{fig:aishell1_et} compares the trade-off between accuracy and latency for the proposed model with ET or without ET, on AISHELL-1, which indicates a better trade-off can be obtained with ET.


\textbf{Comparing with other methods.} Overall, the proposed model achieves competitive ASR performance. For example, on the AISHELL-1 dataset, the CER and latency of the proposed model are both noticeably lower than the well-established and widely-used chunk Conformer CTC model, i.e. 5.55 versus 7.49 of CER, and 494 ms versus 727 ms of latency. The CER of the proposed model is also much lower than other systems shown in the first part of Table~\ref{tab:aishell-1}. On AISHELL-2, the proposed model achieves comparable CER with the advanced CIF model, i.e. 6.08 versus 6.04. But the latency of the proposed model, i.e. 764 ms, is much lower than the 2.56 s chunk used in the CIF model.

\section{Conclusions}
\label{sec:conclusion}
This work explores the efficiency of Mamba for streaming ASR within the UMA framework. Our experiments show that the recursive nature of Mamba is especially suitable for streaming speech learning. The UMA framework fits well with streaming ASR by detecting token activities with unimodal weights. As a whole, the proposed streaming ASR model achieves superior ASR performance in terms of both recognition accuracy and latency. 

\vfill\pagebreak

\newpage
\bibliographystyle{IEEEbib}
\bibliography{strings,refs}

\end{document}